\documentclass{ws-procs9x6}%
\usepackage{makeidx}
\usepackage{amsmath}
\usepackage{graphicx}%
\usepackage{amsfonts}%
\usepackage{amssymb}

\begin{document}

\title{QCD GLUEBALL SUM RULES AND VACUUM TOPOLOGY}
\author{HILMAR FORKEL$^{\ast}$}

\address{Institute for Theoretical Physics, University of Heidelberg \\
D-69102 Heidelberg, Germany  \\ and \\ IFT-UNESP \\
01405-900 - S\~{a}o Paulo, SP, Brazil\\ 
$^{\ast}$E-mail: forkel@ift.unesp.br} 

\begin{abstract}
Several key problems of QCD sum rules in the spin-0 glueball channels are
resolved by implementing nonperturbative short-distance physics from direct
instantons and topological charge screening. A lattice-based instanton size
distribution and the IR renormalization of the nonperturbative Wilson
coefficients are also introduced. Results of a comprehensive quantitative sum
rule analysis are reviewed and their implications discussed.

\end{abstract}

\bodymatter           

\section{Introduction}

The gluonium states of QCD have remained intriguing for almost four
decades\cite{gel72}. Their ``exotic'' nature reflects itself not least in
several long-standing problems which the QCD sum rule approach faces in the
spin-0 glueball channels\cite{nar98}. In the scalar ($0^{++}$) glueball
correlator, in particular, the departure from asymptotic freedom sets in at
unusually small distances\cite{nov81} and the perturbative Wilson coefficients
of the standard operator product expansion (OPE) proved inadequate to
establish consistency both among the $0^{++}$ glueball sum rules and with an
underlying low-energy theorem\cite{for01}. Although nonperturbative
contributions due to direct (i.e. small) instantons\cite{sch98} were early
candidates for the missing short-distance physics\cite{nov280}, insufficient
knowledge of the instanton size distribution\cite{sch98,rin99} prevented their
quantitative implementation at the time. Only recently, the derivation of the
exact instanton contributions (to leading order in $\hbar$), their duality
continua and the corresponding Borel sum-rule analysis\cite{for01} showed that
direct instantons indeed solve the mentioned key problems in the scalar
glueball channel. Below we will outline a more thorough and systematic
treatment\cite{for03} which eliminates artefacts of earlier approximations,
significantly modifies the sum-rule results in both spin-0 channels and
improves their reliability. Moreover, we implement for the first time
topological charge screening contributions to the OPE in the $0^{-+}$
channel\cite{for03} and show how those resolve earlier sum-rule
inconsistencies\cite{zha03}. Finally, we review several new predictions for
spin-0 glueball properties.

\section{Correlators and sum rules}

The spin-0 glueball sum rules are based on the scalar $\left(  0^{++}\right)
$ and pseudoscalar $\left(  0^{-+}\right)  $ correlations functions
\begin{equation}
\Pi_{G}(-q^{2})=i\int d^{4}x\,e^{iqx}\left\langle 0|T\,O_{G}\left(  x\right)
O_{G}\left(  0\right)  |0\right\rangle \label{corr}%
\end{equation}
where $O_{G}$ with $G\in\left\{  S,P\right\}  $ are the gluonic interpolating
fields (of lowest mass dimension) $O_{S}=\alpha_{s}G_{\mu\nu}^{a}G^{a\mu\nu}$
and $O_{P}=\alpha_{s}G_{\mu\nu}^{a}\tilde{G}^{a\mu\nu}$ with $\tilde{G}_{\mu\nu
}\equiv\left(  1/2\right)  \varepsilon_{\mu\nu\rho\sigma}G_{\rho\sigma}$ (the
definition in Ref. \refcite{for03} contains a typo). The zero-momentum limit of
the correlator (\ref{corr}) is\ governed by the low-energy theorems (LETs)
\begin{equation}
\Pi_{S}\left(  q^{2}=0\right)  =\frac{32\pi}{b_{0}}\left\langle \alpha
G^{2}\right\rangle \label{sLET}%
\end{equation}
in the scalar\cite{let} and (for three light flavors and $m_{u,d}\ll m_{s}$)%
\begin{equation}
\Pi_{P}\left(  q^{2}=0\right)  =\left(  8\pi\right)  ^{2}\frac{m_{u}m_{d}%
}{m_{u}+m_{d}}\left\langle \bar{q}q\right\rangle \label{psLET}%
\end{equation}
in the pseudoscalar channel\cite{leu92}. (Note that Eq. (\ref{psLET}) vanishes
in the chiral limit.) Consistency with the low-energy theorems places
stringent constraints on the sum rules which cannot be satisfied without
nonperturbative short-distance physics\cite{for01,for03}.

Contact with the hadronic information in the glueball correlators is
established by means of the dispersive representation%
\begin{equation}
\Pi_{G}\left(  Q^{2}\right)  =\frac{1}{\pi}\int_{0}^{\infty}%
ds\frac{\operatorname{Im}\Pi_{G}\left(  -s\right)  }{s+Q^{2}} \label{disprel}%
\end{equation}
where the necessary number of subtractions is implied but not written
explicitly. The standard sum-rule description of the spectral functions
\begin{equation}
\operatorname{Im}\Pi_{G}^{\left(  ph\right)  }\left(  s\right)
=\operatorname{Im}\Pi_{G}^{\left(  pole\right)  }\left(  s\right)
+\operatorname{Im}\Pi_{G}^{\left(  cont\right)  }\left(  s\right)
\label{phspecdens}%
\end{equation}
contains one or two resonance poles in zero-width approximation,
$\operatorname{Im}\Pi^{\left(  pole\right)  }\left(  s\right)  =\pi\sum
_{i=1}^{2}f_{Gi}^{2}m_{Gi}^{4}\delta\left(  s-m_{Gi}^{2}\right)  $, and the
local-duality continuum $\operatorname{Im}\Pi_{G}^{\left(  cont\right)
}\left(  s\right)  =\theta\left(  s-s_{0}\right)  \operatorname{Im}\Pi
_{G}^{\left(  IOPE\right)  }\left(  s\right)  $ from the IOPE discontinuities
in the ''duality range'' which starts at an effective threshold $s_{0}$.

In order to write down QCD sum rules, the Borel-transformed IOPE and
dispersive representations of the correlators - weighted by $k$ powers of
$-Q^{2}$ - are matched in the fiducial $\tau$-region ($\tau$ is the Borel
variable) and rearranged in terms of the continuum-subtracted Borel moments
$\mathcal{R}_{G,k}$ as
\begin{align}
\mathcal{R}_{G,k}\left(  \tau;s_{0}\right)   &  \equiv\frac{1}{\pi}\int
_{0}^{s_{0}}dss^{k}\operatorname{Im}\Pi_{G}^{\left(  IOPE\right)  }\left(
s\right)  e^{-s\tau}\label{csubtrbmoms}\\
&  =\sum_{i=1}^{2}f_{Gi}^{2}m_{Gi}^{4+2k}e^{-m_{Gi}^{2}\tau}-\delta_{k,-1}%
\Pi_{G}^{\left(  ph\right)  }(0).\nonumber
\end{align}
The pole contributions of interest (and for $k=-1$ the crucial subtraction
terms) are then isolated on the RHS, and the hadronic parameters
$m_{Gi},f_{Gi}$ and $s_{0}$ can be determined numerically.

\section{IOPE}

Our theoretical framework, the instanton-improved operator product expansion
(IOPE), factorizes the correlators at large, spacelike momenta $Q^{2}%
\equiv-q^{2}\gg\Lambda_{QCD}$ into contributions from ``hard'' field modes
(with momenta $\left|  k\right|  >\mu$) in the Wilson coefficients $\tilde
{C}_{D}\left(  Q^{2}\right)  $ and ``soft'' field modes (with $\left|
k\right|  \leq\mu$) in the ``condensates'' $\left\langle \hat{O}_{D}%
\right\rangle _{\mu}$ of operators $\hat{O}_{D}$ with increasing dimension
$D$. Previous glueball sum rules based on the OPE with purely perturbative
Wilson coefficients were plagued by notorious inconsistencies between the
predictions of different moment sum rules and by massive LET violations.
Moreover, the soft nonperturbative condensate contributions were exceptionally small.

We have therefore analyzed \emph{hard} nonperturbative contributions to the
Wilson coefficients. They are strongly channel dependent and due to direct
instantons and topological charge screening. The instanton contributions to
the spin-0 coefficients\cite{for03},
\begin{equation}
\Pi_{G}^{\left(  I+\bar{I}\right)  }\left(  x^{2}\right)  =\frac{2^{8}3}%
{7}\int d\rho n\left(  \rho\right)  \frac{1}{\rho^{4}}\,_{2}F_{1}\left(
4,6,\frac{9}{2},-\frac{x^{2}}{4\rho^{2}}\right)  , \label{pix}%
\end{equation}
are large (while those in the $2^{++}$ tensor channel vanish) and add to the
unit-operator coefficients $\tilde{C}_{0}^{\left(  G\right)  }$. The imginary
part of their Fourier transform at timelike momenta generates the Borel
moments\cite{for01}%
\begin{align}
\mathcal{R}_{k}^{\left(  I+\bar{I}\right)  }\left(  \tau\right)   &
=-2^{7}\pi^{2}\delta_{k,-1}\int d\rho n\left(  \rho\right)  -2^{4}\pi^{3}\int
d\rho\\
&  \times n\left(  \rho\right)  \rho^{4}\int_{0}^{s_{0}}dss^{k+2}J_{2}\left(
\sqrt{s}\rho\right)  Y_{2}\left(  \sqrt{s}\rho\right)  e^{-s\tau}\nonumber
\end{align}
which are similar or larger in size than the perturbative ones. The evaluation
of these moments requires as the sole input the (anti-) instanton distribution
$n\left(  \rho\right)  $ which is implemented by means of a lattice-based
Gaussian-tail parametrization with the correct small-$\rho$ behavior,
\begin{equation}
n_{g}\left(  \rho\right)  =\frac{2^{18}}{3^{6}\pi^{3}}\frac{\bar{n}}{\bar
{\rho}}\left(  \frac{\rho}{\bar{\rho}}\right)  ^{4}\exp\left(  -\frac{2^{6}%
}{3^{2}\pi}\frac{\rho^{2}}{\bar{\rho}^{2}}\right)
\end{equation}
(for $N_{c}=N_{f}=3$), which was introduced in Ref. \refcite{for03}  and shown
to prevent several artefacts of the oversimplified ``spike'' approximation
$n(\rho)=\bar{n}\delta\left(  \rho-\bar{\rho}\right)  $ on which all previous
direct-instanton calculations had relied.

As an additional benefit, the realistic size distribution allows for a
gauge-invariant IR renormalization which excludes large instantons with size
$\rho>\mu^{-1}$ from the Wilson coefficients,
\begin{equation}
n\left(  \rho\right)  \rightarrow\tilde{n}\left(  \mu;\rho\right)
\equiv\theta_{\beta}\left(  \rho-\mu^{-1}\right)  n\left(  \rho\right)
\end{equation}
($\theta_{\beta}$ is a soft step function). The instanton-induced $\mu$
dependence turns out to be relatively weak for $\mu<\bar{\rho}^{-1}$, as
necessary to compensate its perturbative counterpart. Neglect of this
renormalization, although common practice in perturbative Wilson coefficients,
would significantly contaminate the results, e.g. by missing the reduction of
the direct-instanton density%
\begin{equation}
\bar{n}=\int_{0}^{\infty}d\rho n\left(  \rho\right)  \rightarrow\int
_{0}^{\infty}d\rho\tilde{n}\left(  \mu;\rho\right)  \equiv\bar{n}\left(
\mu\right)  .
\end{equation}
Another important renormalization effect is the reduction of the instanton
contributions to the pseudoscalar relative to the scalar sum rules.

The instanton's self-duality causes a strongly repulsive contribution to the
$0^{-+}$ channel, with seemingly detrimental impact on the sum
rules\cite{zha03}: the glueball signal disappears and both unitarity and the
LET (\ref{psLET}) are badly violated. The origin of these problems can be
traced to the neglect of topological charge screening\cite{for03}. Due to
their high channel selectivity and a small screening length $\lambda_{D}\sim
m_{\eta^{\prime}}^{-1}\sim0.2$ fm, the model-independent screening
correlations\cite{dow92}
\begin{equation}
\Pi_{P}^{\left(  scr\right)  }\left(  x\right)  \simeq-2^{8}\pi^{2}\left(
\xi\gamma_{\eta_{0}}\right)  ^{2}\left\langle \eta_{0}\left(  x\right)
\eta_{0}\left(  0\right)  \right\rangle
\end{equation}
($\xi$ is the overall topological charge density ($=\bar{n}$ for instantons)
and $\eta_{0}$ the flavor-singlet part of the $\eta^{\prime}$) affect almost
exclusively the $0^{-+}$ Wilson coefficients. They arise from the
axial-anomaly induced attractive (repulsive) interaction between topological
charge lumps of opposite (equal) sign due to $\eta_{0}$ exchange\cite{div80}
and, after correcting for $\eta_{0}-\eta_{8}$ mixing, add the terms
\begin{equation}
\mathcal{R}_{P,k}^{\left(  scr\right)  }\left(  \tau\right)  =-\delta
_{k,-1}\left(  \frac{F_{\eta^{\prime}}^{2}}{m_{\eta^{\prime}}^{2}%
}+\frac{F_{\eta}^{2}}{m_{\eta}^{2}}\right)  +F_{\eta^{\prime}}^{2}%
m_{\eta^{\prime}}^{2k}e^{-m_{\eta^{\prime}}^{2}\tau}+F_{\eta}^{2}m_{\eta}%
^{2k}e^{-m_{\eta}^{2}\tau} \label{rscr}%
\end{equation}
to the pseudoscalar IOPE moments.

They strongly reduce the direct-instanton induced repulsion and resolve the
disastrous problems mentioned above\cite{for03}: positivity of the spectral
function is restored, the four $0^{-+}$ Borel sum rules (\ref{csubtrbmoms})
are stable and contain consistent pseudoscalar glueball information. (All
previous analyses had discarded the $k=-1$ sum rule and thereby missed
valuable first-principle information and LET consistency checks.)

\section{Results and discussion}

We have implemented direct instanton and topological charge screening
contributions into the OPE coefficients of the spin-0 glueball correlators,
evaluated their duality continuua and demonstrated how these contributions
resolve the problems of previous QCD sum rule analyses\cite{for03}. A
lattice-based instanton size distribution and the gauge-invariant IR
renormalization of the nonperturbative Wilson coefficients were also
introduced. Quark admixtures, and thereby quarkonium mixing effects, enter
through quark loops, the instanton size distribution and the condensates.

In the scalar channel, the sizeable direct instanton contributions are
indispensable for mutually and LET consistent sum rules. Their improved
treatment reduces our earlier (spike-distribution based) result for the
$0^{++}$ glueball mass to $m_{S}=1.25\pm0.2$ GeV. (The mass stays well beyond
1 GeV, however, in contrast to obsolete predictions based on purely
perturbative coefficients.) This value is somewhat smaller than the quenched
lattice results\cite{lat} (which will probably be reduced by light-quark
effects) and consistent with the broad glueball state found in a comprehensive
$K$-matrix analysis\cite{ani03}. The systematics among our different Borel
moments likewise indicates a rather large width of the scalar glueball,
$\Gamma_{S}\gtrsim0.3$ GeV.

Our prediction for the scalar glueball decay constant, \ $f_{S}=1.05\pm0.1$
GeV, is several times larger than the value obtained when ignoring the
nonperturbative Wilson coefficients. This implies an exceptionally small
$0^{++}$ glueball size, in agreement with several lattice results\cite{ish02}.
Another stringent, OPE- and sum-rule-independent consistency check of the
instanton contributions and their $f_{S}$ enhancement provide numerical
simulations in an instanton ensemble\cite{sch95} (ILM). For $\left|  x\right|
\lesssim0.5$ fm the scalar ILM and IOPE correlators are indeed very similar,
and the ILM prediction $f_{S}^{(ILM)}=0.8$ GeV is similarly large. This
indicates a robust instanton effect and rules out that the large\ $f_{S}$
''may signal some eventual internal inconsistencies in the treatment of the
instanton contributions''\cite{nar05}. A subsequent model-independent
confirmation of the $f_{S}$ enhancement was supplied by the first direct
(quenched) lattice result\cite{che05} $f_{S}=0.86\pm0.18$ GeV which is
consistent both with our prediction and the ILM value (the latter is
practically independent of quark quenching\cite{sch95}).

Our prediction for $f_{S}$ implies enhanced partial widths for radiative
$J/\psi$ and $\Upsilon$ decays into scalar glueballs and is therefore relevant
for experimental glueball searches, e.g. in the CLEO and BES data on
$\Upsilon\rightarrow\gamma f_{0}$ and other decay branches. Since the
exceptionally small size and large decay constant are particular to the
\emph{scalar} glueball and since only part of the scalar\ decay constant
contributes to the radiative production rates\cite{he02}, however, the above
results are not ruled out by experimental data on decays into $0^{-+}$
glueballs (cf. Ref. \refcite{nar05}).

In the pseudoscalar ($0^{-+}$) glueball correlator we have identified and
implemented a new type of nonperturbative contributions to the Wilson
coefficients, due to topological charge screening. Roughly speaking, the
screening effects ``unquench'' the direct instanton contributions, thereby
restoring unitarity, the axial Ward identity and the resonance signals.
Consistency among all moment sum rules and with the underlying LET is also
achieved, and the resulting mass prediction $m_{P}=2.2\pm0.2$ GeV lies inside
the range of quenched and unquenched lattice data. The coupling $f_{P}%
=0.6\pm0.25$ GeV is somewhat enhanced by the topological short-distance
physics, affecting radiative production rates, the $\gamma\gamma\rightarrow
G_{P}\pi^{0}$ cross section at high momentum transfers and other glueball signatures.

The crucial impact of the nonperturbative Wilson coefficients on both spin-0
glueball correlators is particularly evident in the interplay between their
subtraction constants. Indeed, the notorious consistency problems which
plagued previous $0^{++}$ glueball sum rules were primarily caused by the
large LET-induced subtraction constant $\Pi_{S}\left(  0\right)  \simeq0.6$
GeV$^{4}$ (cf. Eq. (\ref{sLET})): it cannot be matched by perturbative Wilson
coefficients (since $\Pi_{S}^{\left(  pert\right)  }\left(  0\right)  =0$) and
requires the direct instanton contribution $\Pi_{S/P}^{\left(
I,\bar{I}\right)  }\left(  0\right)  =\pm2^{7}\pi^{2}\bar{n}_{dir}\simeq0.63$
GeV$^{4}$. At first sight this seems to imply a conflict with the much smaller
LET subtraction constant $\Pi_{P}\left(  0\right)  \simeq-0.02$ GeV$^{4}$ (cf.
Eq. (\ref{psLET})) in the $0^{+-}$ channel, however, since the instanton
contributions to both spin-0 correlators are equal (up to a sign). Here the
topological screening contributions $\Pi_{P}^{\left(  scr\right)  }\left(
0\right)  \simeq0.59$ GeV$^{4}$ from Eq. (\ref{rscr}) prove indispensable:
they restore consistency by canceling most (and in the chiral limit, where
$\Pi_{P}\left(  0\right)  \rightarrow0$, all) of the instanton contributions:
$\Pi_{P}^{\left(  scr\right)  }\left(  0\right)  +\Pi_{P}^{\left(
I,\bar{I}\right)  }\left(  0\right)  \simeq\Pi_{P}\left(  0\right)  $.

To summarize: contrary to naive expectation, the nonperturbative contributions
to the OPE\ of the spin-0 glueball correlators reside primarily in the Wilson
coefficients (i.e not in the condensates) and are closely related to the
topological vacuum structure. The nonperturbative short-distance physics
resolves long-standing consistency problems of the associated QCD sum rules
and generates a rather diverse set of new glueball predictions. A large part
of the $0^{++}$ glueball mass and binding originates from direct instantons,
for example, while their net effects in the $0^{+-}$ channel are smaller and
more subtle, due to cancellations between instanton and topological charge
screening contributions.

This work was supported by FAPESP and CNPq of Brazil.

\end{document}